\documentclass{article}
\usepackage{graphicx}
\usepackage{bm}
\usepackage{amssymb}
\usepackage{amsmath}
\usepackage{amsfonts}
\usepackage[usenames]{color}
\usepackage{hyperref}
\usepackage[english]{babel}
\usepackage[T2A]{fontenc}
\newcommand\nnfootnote[1]{%
  \begin{NoHyper}
  \renewcommand\thefootnote{}\footnote{#1}%
  \addtocounter{footnote}{-1}%
  \end{NoHyper}
}

\begin{document}

\title{Quantifying the transition from spiral waves to spiral wave chimeras in a lattice of self-sustained oscillators}

\author{I.A. Shepelev \footnotemark[1], A.V. Bukh \footnotemark[1], S.S. Muni \footnotemark[2], V.S. Anishchenko\footnotemark[1]}

{\renewcommand\thefootnote{\fnsymbol{footnote}}%
  \footnotetext[1]{Department of Physics, Saratov State University, 83 Astrakhanskaya Street, Saratov, 410012, Russia}\footnotetext[2]{School of Fundamental Sciences, Massey University, Palmerston North, New Zealand}}
\maketitle
\begin{abstract}
 The present work is devoted to the detailed quantifying the transition 
 from spiral waves to spiral wave chimeras in a network of self-sustained oscillators with two-dimensional geometry. The basic elements of the networks are the van der Pol oscillator and the FitzHugh-Nagumo neuron. The both models are in the regime of relaxation oscillations. We analyze the regime with using the indexes of local sensitivity which enables us to evaluate the sensitivity of each individual oscillator at finite time. Spiral waves are observed in the both lattices when the interaction between elements have the local character. The dynamics of all the elements is regular. There are no any expressed high-sensitive regions. We have discovered that when the coupling becomes nonlocal the features of the systems are significantly changed. The oscillation regime of the spiral wave center element switches to chaotic one. Beside this, a region with the high sensitivity occurs around this oscillator. Moreover, we show that the latter expands in space with elongation of the coupling range. As a result, an incoherence cluster of the spiral wave chimera is formed exactly within this high-sensitive area. Forming this cluster is accompanied by the sharp increase in values of the maximal Lyapunov exponent to the positive region. Furthermore, we explore that the system can even switch to hyperchaotic regime, when several Lyapunov exponents becomes positive.

\end{abstract}


\nnfootnote{Keywords: Spatiotemporal patterns, chimera, van der Pol oscillator, FitzHugh-nagumo neuron, spiral waves, spiral wave chimera, Nonlocal Interactions, Lyapunov exponent}
\nnfootnote{E-mail addresses: I.A. Shepelev (\url{igor\_sar@li.ru}), A.V. Bukh (\url{buh.andrey@yandex.ru}), S.S. Muni(\url{s.muni@massey.ac.nz}), V.S. Anishchenko (\url{wadim@info.sgu.ru})}

\section*{\label{sec:intro}Introduction}

Investigation of auto-wave processes in nonlinear complex systems and their  mathematical models is one of the highly developed direction in nonlinear physics and related fields. The auto-wave regime is the self-sustained wave regime in a non-equilibrium system which weakly depends on initial and boundary conditions. These systems can be described by a system of ordinary differential equations with diffusive coupling and an active nonlinearity \cite{kopell1973plane,winfree1974rotating,hagan1982spiral, keener1986spiral,vasiliev1987autowave, biktashev1989diffusion, barkley1990spiralwave, ermakova2005blood}. Intensive studies of complex spatiotemporal behavior in such models has begun from the 50s of XX century and continues to this day \cite{kuramoto2003rotating, zhang2004spiral, ermakova2005blood, martens2010solvable, kuzmin2012deformation, tang2014novel, panaggio2015Bchimera, xie2015twisted, li2016spiral, weiss2017weakly, totz2018spiral}.  
The interest to the study of autowave structures in different nonlinear complex systems  sharply increased in the early 2000s. This has been mainly caused by the discovery of so-called ''chimera states'' which are hybrid states that emerge spontaneously, combining both coherent and incoherent parts. The first chimera state has been found in a ring of non-locally coupled identical phase oscillators by Kuramoto and Battogtokh in \cite{kuramoto2002coexistence} in 2002. The term ''chimera state'' has been proposed in \cite{abrams2004chimera} in 2004 by Abrams and Strogatz. The non-locality of coupling means  that each individual oscillator is symmetrically coupled with all adjacent oscillators from the neighborhood with a certain radius \cite{tanaka2003complex}. 
Later these states have been found in diverse oscillatory models with nonlocal interaction such as the FitzHugh-Nagumo, the van der Pol, the Stuart-Landau, the Hindmarsh-Rose and others \cite{tinsley2012chimera, panaggio2015Bchimera,martens2013chimera, rosin2014transient, schmidt2014coexistence,kapitaniak2014imperfect, omelchenko2011loss, Maistrenko-2015, Zakharova-2014,shepelev2018chimera}. 
However, the chimera structures have been discovered in systems with purely local coupling \cite{laing2015chimeras, clerc2016chimera,tang2014novel,li2016spiral,shepelev2018local} and with global one \cite{yeldesbay2014chimeralike, schmidt2014coexistence}.
The chimeras have been found in real experiments \cite{hagerstrom2012experimental,tinsley2012chimera,viktorov2014coherence}. 
The potential relevance of chimera states also includes  the phenomenon of unihemispheric sleep which is observed in birds and dolphins \cite{rattenborg2000behavioral}. They sleep with only one eye open, meaning that half of the brain is synchronous with the other half being asynchronous. Furthermore, it is hypothesized that chimera states are the route of both onset and termination of epileptic seizures \cite{mormann2003epileptic,rothkegel2014irregular,andrzejak2016}. Chimera can be realized even in the common brain dynamics as it has been shown in the recent work \cite{bansal2019cognitive}.

The investigation on the lattice dynamics of coupled oscillators has revealed a special type of spiral wave, namely spiral wave chimeras (SWC), first found numerically in two-dimensional nonlocally coupled systems \cite{kuramoto2003rotating,shima2004rotating} and confirmed analytically for the 2D system of  nonlocally coupled phase oscillators \cite{martens2010solvable,laing2017chimeras}. The spiral wave chimera has been observed experimentally in \cite{totz2018spiral}  in catalyst-free Belousov–Zhabotinsky reaction. SWC has been discovered in different 2D systems \cite{omelchenko2012stationary,tang2014novel,guo2018spiral,schmidt2017chimera,bukh2019bspiral,shepelev2019variety}. Usually the elongation of the coupling range leads to the transformation of the common spiral wave to the spiral wave chimera, as well as in the other models. 
However, despite the extensive numerical evidence of SWC, their confirmation in many dynamical systems and the experimental observations, the processes that occur during the formation of SWC remains elusive.
In our previous work \cite{shepelev2019variety}, we had offered a hypothesis that the reason behind the formation of  SWC in the lattice of  bistable FHN oscillators is in strong increase of the sensitivity of a region around the wave center when the coupling range $r $ is elongated. We try to show that this phenomenon is common in the oscillatory models, where the SWC are observed, namely lattices of the van der Pol oscillators and of the FitzHugh-Nagumo oscillators. The interaction between oscillators has linear nonlocal character. Besides, we show the similar results for a lattice of discrete-time elements (Nekorkin maps).

\section{\label{sec:system}The models}

In this section, we summarize two models considered in this article, namely — the van der Pol (vdP) and FitzHugh-Nagumo (FHN) oscillators. We  also introduce the respective coupling schemes describing the two-dimensional layout with nonlocal interactions.

\subsection{Lattice of van der Pol oscillators}

The dynamics of a single van der Pol oscillator is described by the following system of ODEs:
\begin{equation}
\begin{array}{l}
\dfrac{dx}{dt}  = y,\\
\dfrac{dy}{dt} = \mu(1-x^2)y - \omega^2 x,
\end{array}
\label{eq:vdP_single}
\end{equation}
where $x $ and $y $ are dynamical variables. The parameter $\mu$ determines the nonlinearity degree, while the parameter $\omega$ is responsible for the oscillator frequency.  The value of $\mu=0$ corresponds to the supercritical Andronov-Hopf bifurcation which is results in the birth of limit cycle for $\mu>0$. The Values of control parameters are fixed as $\mu=2.1 $ and $\omega=2.5 $ in this study. These values correspond to the relaxation regime of self-oscillations.

We consider the model of a spatially organized  ensemble of oscillators which is a 2D regular  $N \times N $  lattice with an edge $N=100 $ and consists of nonlocally coupled vdP oscillators \eqref{eq:vdP_single}. The interaction between elements is introduced linearly into $x $-variables.  This model is described by the following system of network equations:
%
\begin{equation}
\begin{array}{l}
\dfrac {dx_{i,j}}{dt} = y_{i,j}  + \dfrac {\sigma} {Q_{i, j}}
\sum\limits_{\tiny {k,p} }
\left(x_{k,p} - x_{i,j}\right),
\\
\dfrac{dy_{i,j}}{dt} = \mu (1- x_{i,j}^2) y_{i,j} - \omega^2x_{i,j} ,\\[8pt]
i,j=1,...N,
\end{array}
\label{eq:vdP_grid}
\end{equation}
%
The double index of the dynamic variables $ x_{i, j} $ and $ y_{i, j} $ with $ i,j = 1, ..., N $ determines the position of an element in the two-dimensional lattice. All the oscillators are identical in parameters and each of them is coupled with all the lattice elements from a square with side $ (1 + 2P) $ in the center of which this element is located. The integer $ P $ defines the nonlocality of coupling and is called the interaction interval.  The case of $P=1 $ corresponds to the local coupling, while $P=N/2 $ is the case of global coupling, when each element interacts with the whole system. 
The number of coupled units with the $i,j $th element is determined by the parameter $Q_{i, j}$, which is given by a number of $k,p$ combinations with the following relations:
\begin{equation}\label{eq:vdP_boundaries}
\left\{\begin{aligned}
&\max(1,i-P) \leqslant k,p \leqslant \min(N,i+P), \\
&\max(1,j-P) \leqslant k,p \leqslant \min(N,j+P),
\end{aligned}\right.
\end{equation}
which represent zero flux boundary conditions (Neumann type) in the nonlocal case~\cite{haugland2015self}. We also use the notion of the coupling range $r=P/N $ in analogy with the classical works on chimeras. The coupling strength is determined by the value of $\sigma $, which is fixed in this study as $\sigma=0.9 $.

\subsection{FitzHugh-Nagumo model}

The FHN oscillator is one of the simplest model which describes the neuron dynamics. The behavior of a single oscillator is described by following system of differential equations:
\begin{equation}
\begin{array}{l}
\varepsilon \dfrac{du}{dt} = u - \alpha u^3 -w,\\
\dfrac{dw}{dt} = \gamma v - w + \beta,
\end{array}
\label{eq:FHN_single}
\end{equation}
where $u $ and $w $ denote a fast activator and a slow inhibitor variable, respectively. The parameter $\varepsilon $ determines the time-scale separation and is fixed in this study at $\varepsilon=0.2 $. The threshold parameters $\beta $ and $\gamma $ determines the oscillatory, excitable, or bistable behavior in the system, i.e., one (self-sustained or excitable regime) or three (bistable regime) times it intersecs the nullclines. We fix the value of $\beta=0.001 $ throughout this study just to avoid the full symmetry of the system. Hence, when $\gamma < 0.72 $ the regime is bistable with two stable foci and a saddle between them (bistable regime). The value of $\gamma \approx 0.72 $ corresponds to the subcritical Andronov-Hopf bifurcation. When $\gamma $ exceeds this value, a stable limit cycle appears in the system \eqref{eq:FHN_single} (self-sustained regime). We set the value of $\gamma $ according to the self-sustained regime as $\gamma=0.8 $.

Similar to the case of the vdP lattice \eqref{eq:vdP_grid}, we consider the FHN lattice dynamics on a two-dimensional regular $N \times N $ lattice with $N = 100 $ nodes and zero flux boundary conditions. The FHN coupled lattice dynamics is described as follows:
\begin{equation}
\begin{array}{l}
\varepsilon \dfrac {du_{i,j}}{dt} = u_{i,j} - \alpha u_{i,j}^3 -w_{i,j} + \dfrac {\varepsilon\sigma} {Q}
\sum\limits_{\tiny {k,p} }
\left(u_{k,p} - u_{i,j}\right),
\\
\dfrac{dw_{i,j}}{dt} = \gamma u_{i,j} - w_{i,j} + \beta,\\[8pt]
i,j=1,...N,
\end{array}
\label{eq:FHN_grid}
\end{equation}
The coupling is linear and is introduced in the first dynamical variable $u$. All the oscillators are identical in parameters and each of them is coupled with the coupling strength $\sigma $ with all the lattice elements from a square with side $ (1 + 2P) $ in the center of which this element is located, where $P $ is the coupling interval. Thus, each element is coupled with $Q$ neighbors, which are given by numbers of $k,p $ combinations by formula \eqref{eq:vdP_boundaries} for the zero flux boundary conditions. The coupling strength $\sigma $ is also fixed in this study and is equal to $\sigma=0.8 $.

As initial conditions, we use the instantaneous state of the systems in the regime of a spiral wave for the case of local coupling, which has been obtained for a multitude of random initial values of the variables with a uniform distribution within $x_0 \in [-1,1],~y_0 \in [-1,1] $. When the coupling range $r $ increases, we use the instantaneous states of a corresponding model at the previous step by $r $ as the initial states for the following value of $r $.
The system equations are integrated using the Runge-Kutta 4th order method with time step $dt=0.001$. All the regimes under study are obtained after the transient process of $t_{\rm trans}=10000 $ time units.

\section{Transition from the spiral wave regime to spiral wave chimeras}

\subsection{Van der Pol model: Spiral waves}

\begin{figure}[!ht]
\centering
\parbox[c]{.45\linewidth}{ 
  \includegraphics[width=\linewidth]{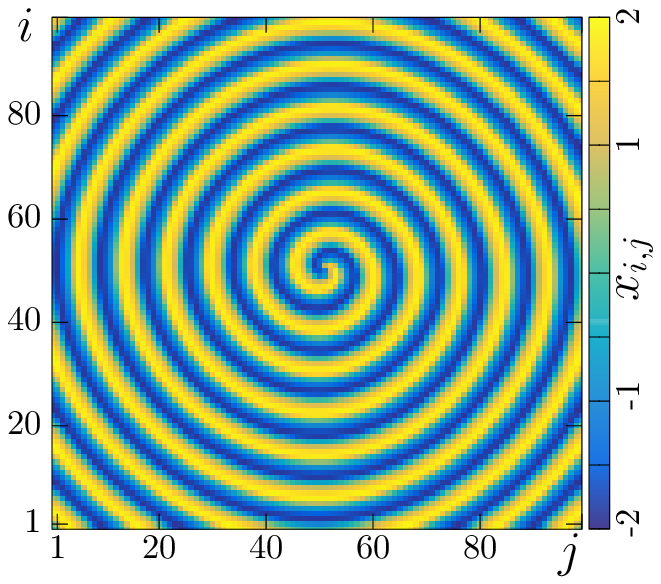}
    \vspace{-9.5mm} \center (a)
\includegraphics[width=\linewidth]{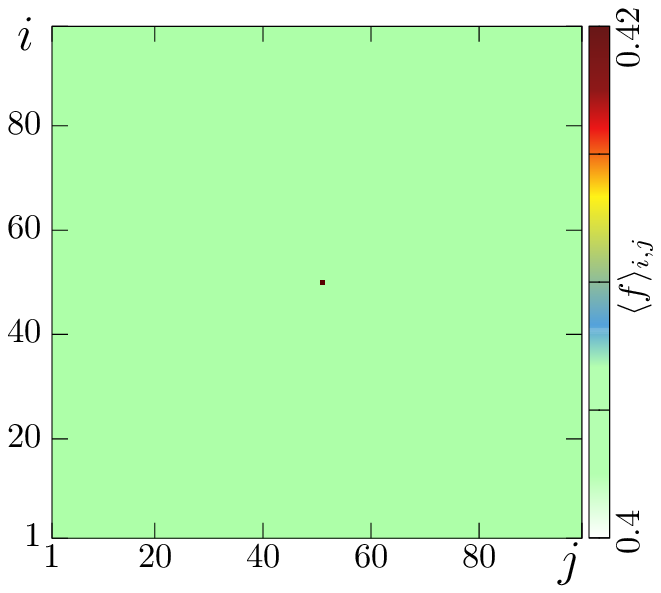}
\vspace{-9.5mm}\center (c) 
}
\parbox[c]{.45\linewidth}{
  \includegraphics[width=\linewidth]{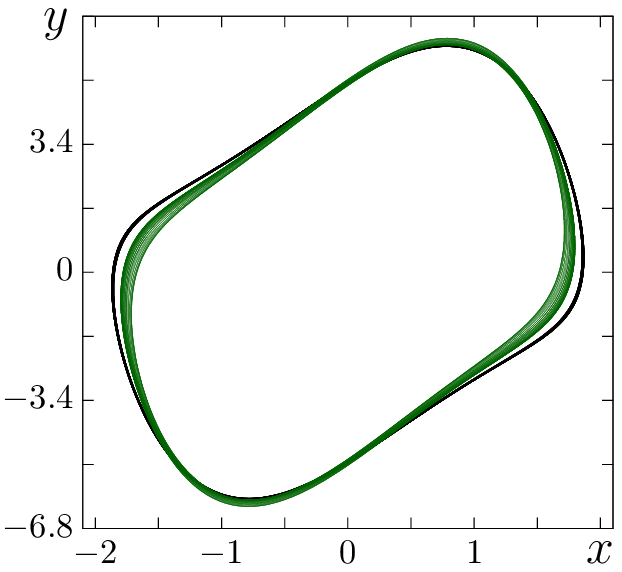}
   \vspace{-9.5mm} \center (b)
\includegraphics[width=\linewidth]{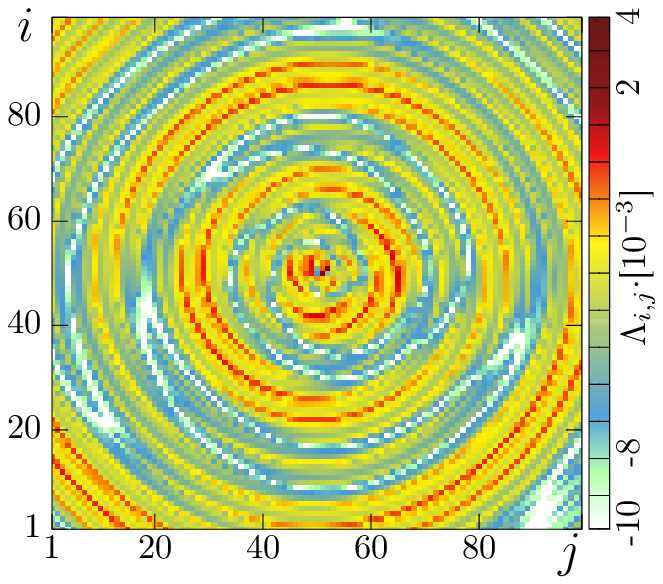}
 \vspace{-9.5mm} \center (d)
}
\caption{(Color online) Spiral wave in \eqref{eq:vdP_grid} for the local coupling $r=0.01$.(a) is a snapshot of the system state, (b) phase portrait projections for oscillators with indexes $i=51,~j=52 $ (wave center oscillator, green line) and  $i=40,~j=40 $(synchronous region, black line), (c) spatial distribution of the mean frequency $f^{i,j}$, (d) spatial distribution of the indexes of local sensitivity (LIS)  $\Lambda_{i,j} $. Parameters: $\sigma=0.9 $, $\mu=2.1 $, $\omega=2.5$, $N=100 $}
\label{fig:vdP_P=1}
\end{figure}

One of the typical regimes realized in \eqref{eq:vdP_grid} for the zero flux BC is a spiral wave \cite{mikhailov2012, pertsov1984rotating, zaritski2002stable}. It's feature is a rotation of the wave front in a spiral fashion around a certain center in space. It is known that the spiral waves are typical for many systems of different nature. They play an important role in living systems \cite{cherry2008visualization, cherry2012mechanisms, bretschneider2009three, pervolaraki2013spatiotemporal}. When the coupling is local ($P=1 $) it is possible to observe a regime with a different number of spiral waves. The wavelength of these waves is  short enough. However, the number of waves can be equal to one too. This case is presented in fig.\ref{fig:vdP_P=1}(a). The wavefront propagates from the center in a spiral fashion. 
Fig.\ref{fig:vdP_P=1}(b) demonstrates phase portrait projections for the wave center (WC) element and the other elements. It is clearly seen that their dynamics are noticeably different. All the lattice elements except  the element of the wave center oscillate periodically, while the WC element oscillates quasi-harmonically, as indicated by the zero value of the maximal Lyapunov exponent ($\lambda_1=0.000493 $).
We calculate the Rice frequency $f^R_{i,j}$ \cite{Rice-1945} for individual $i,j$-oscillators. It is calculated as following:
\begin{equation}
f^{R}_{i,j}=\frac{M_{i,j}}{T},
\label{fR}
\end{equation}
The Rice frequency may be considered as the frequency ($f $) for harmonic oscillations and the mean frequency $ \omega$ for the chaotic or quasi-harmonic ones. 
A spatial distribution of the mean frequency is represented in fig.\ref{fig:vdP_P=1}(c). All the oscillators except the wave center element oscillate with the same frequency. The SW oscillator is characterized by the frequency which is slightly higher than for the rest part of the lattice. 

For the quantitative analysis of the dynamical regimes we plot a spatial distribution of the indixes of local sensitivity (ILS) $\Lambda_{i,j}(T)$, the method of calculation of which has been described in \cite{shepelev2018local}. The ILS evaluates the sensitivity of individual elements to weak perturbations. The ILS can be calculated for each $i $th oscillator by the following formula:
%
\begin{equation}
\Lambda_{i,j}(T):=\dfrac{1}{T}\ln\dfrac{\|\boldsymbol{\xi}_{i,j}(T)\|}{\|\boldsymbol{\xi}_{1,1}(0)\|}=
\dfrac{1}{T}\ln N\dfrac{\|\boldsymbol{\xi}_{i,j}(T)\|}{\|\boldsymbol{\xi}(0)\|},
\label{LIS}
\end{equation}
where $\boldsymbol{\xi}(0) $ is an initial vector of the perturbation of the whole system, while $\boldsymbol{\xi}_{i,j}(T)$  describes the local evolution of a perturbation of the $i, j $th oscillator ($i, j $th component of the perturbation). This characteristic enables us to evaluate the sensitivity of each oscillator of the lattice and to highlight the most unstable spatial regions. The spatial distribution of the ILS for the spiral wave under study is presented in fig.\ref{fig:vdP_P=1}(d). This plot shows that there are no regions with noticeably high sensitivity. High-sensitive regions alternate with low-sensitive ones. 

\subsection{Van der Pol model: Spiral wave chimeras}

\begin{figure}[!ht]
\centering
\parbox[c]{.45\linewidth}{ 
  \includegraphics[width=\linewidth]{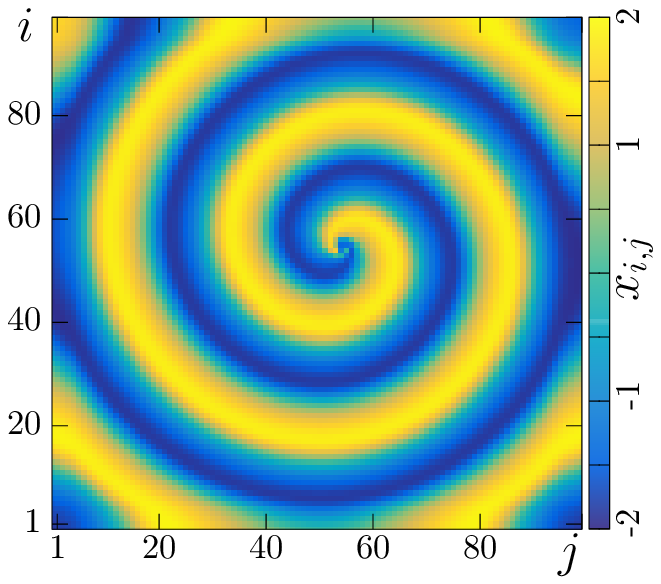}
    \vspace{-9.5mm} \center (a)
\includegraphics[width=\linewidth]{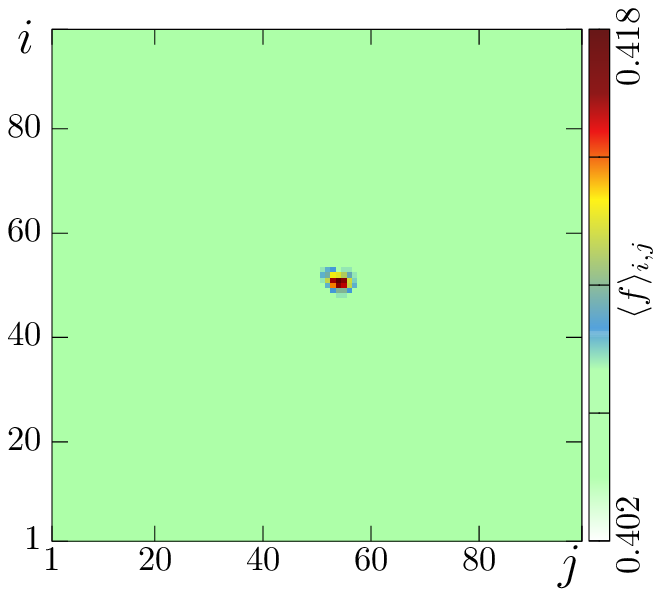}
\vspace{-9.5mm}\center (c) 
}
\parbox[c]{.45\linewidth}{
  \includegraphics[width=\linewidth]{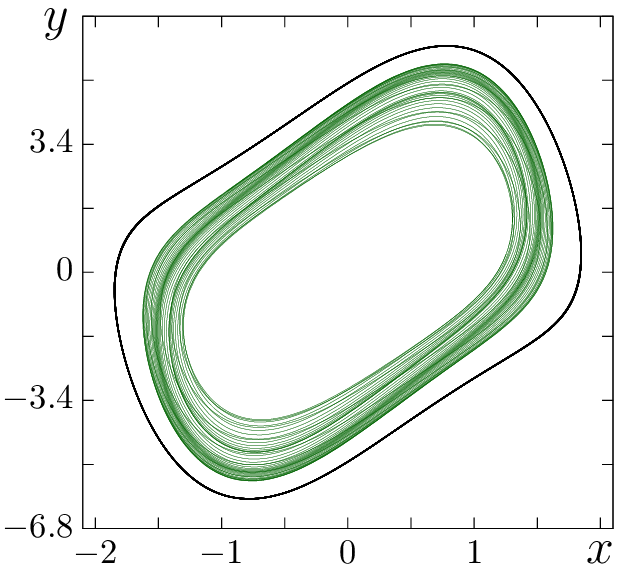}
   \vspace{-9.5mm} \center (b)
\includegraphics[width=\linewidth]{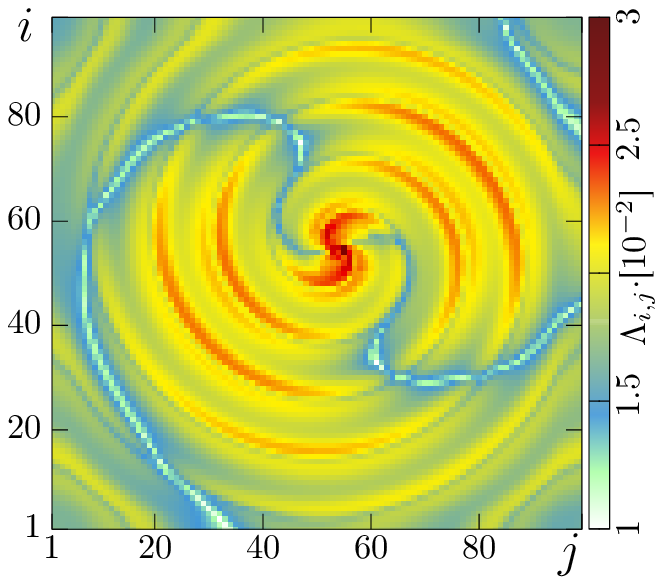}
 \vspace{-9.5mm} \center (d)
}
\caption{(Color online) Spiral wave in \eqref{eq:vdP_grid} for the nonlocal coupling $r=0.04$. (a) is a snapshot of the system state, (b) phase portrait projections for oscillators with indexes $i=55,~j=55 $ (wave center oscillator, green line) and  $i=40,~j=40 $(synchronous region, black line), (c) spatial distribution of the mean frequency $f^{i,j}$, (d) spatial distribution of the LIS  $\Lambda_{i,j} $. Parameters: $\sigma=0.9 $, $\mu=2.1 $, $\omega=2.5$, $N=100 $}
\label{fig:vdP_P=4}
\end{figure}

Now we consider the evolution of the spiral wave regime when the coupling range $r $ increases. At first, we study a wave regime when the value of $r=0.04 $ (small non-locality). An example of the spiral wave for this value of $r $ is shown in fig.\ref{fig:vdP_P=4}(a).
The wavelength becomes noticeably longer than that of the previous case. In the previous case, the transformation to the spiral wave chimera still did not occur and an incoherent core is absent. However, the quantitative analysis of the regime shows that the dynamics has significantly changed. At first, the value of the maximal Lyapunov exponent becomes positive ($\lambda_1=0.023153 $). It means that the chaotic oscillations appear in the system \eqref{eq:vdP_grid}. Fig.\ref{fig:vdP_P=4}(b) shows the phase portrait projections of the wave center element and for the elements outside the wave center. The first phase portrait projection corresponds to a chaotic attractor, and the second one demonstrates the limit cycle of period-1. Thus, oscillations become chaotic only for the elements around the wave center, while oscillations of the other elements remain periodic. As we show below, this behavior is typical for the spiral wave chimera too. A spatial distribution of the mean frequency presented in fig.\ref{fig:vdP_P=4}(c) illustrates that now  a group of oscillators with increased values of the frequency $\langle f \rangle $ forms around the wave center. Moreover, this group already has the bell-like distribution   of $\langle f \rangle $ (i.e. the maximal values of the frequency has the oscillator in the group center) typical for the SWCs. The oscillators outside this group have the same frequency.

A spatial distribution of the ILS in the lattice \eqref{eq:vdP_grid} is shown in fig.\ref{fig:vdP_P=4}(d). It can be concluded from the plot that significant changes occur in the system with an increase in the coupling nonlocality. In the previous case of $r=0.01 $, the oscillators in the wave centers are characterized by the similar values of $\Lambda_{i,j}$, the ones outside the center, i.e. they have been low-sensitive and have not been distinguished from the other oscillators of the wavefront. When $r=0.04 $, oscillators of the wave center become the most sensitive elements of the whole lattice and are characterized by the maximal values of the ILS. Consequently, this part of the system is most likely to develop instability and incoherence.
 
\begin{figure}[!ht]
\centering
\parbox[c]{.45\linewidth}{ 
  \includegraphics[width=\linewidth]{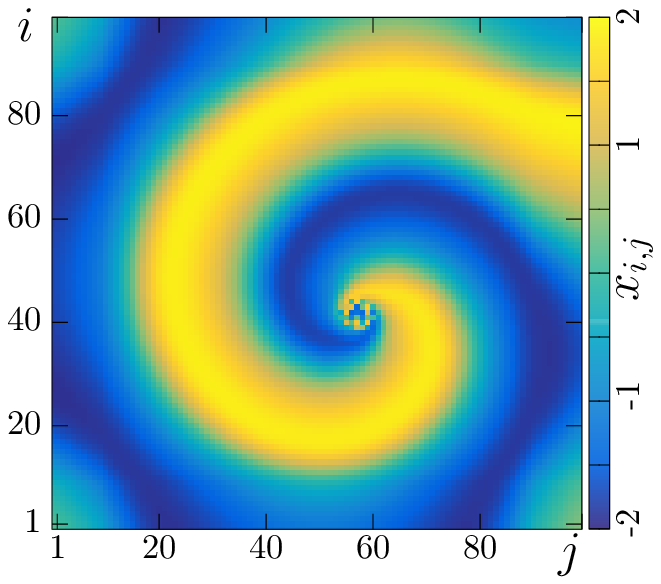}
    \vspace{-9.5mm} \center (a)
\includegraphics[width=\linewidth]{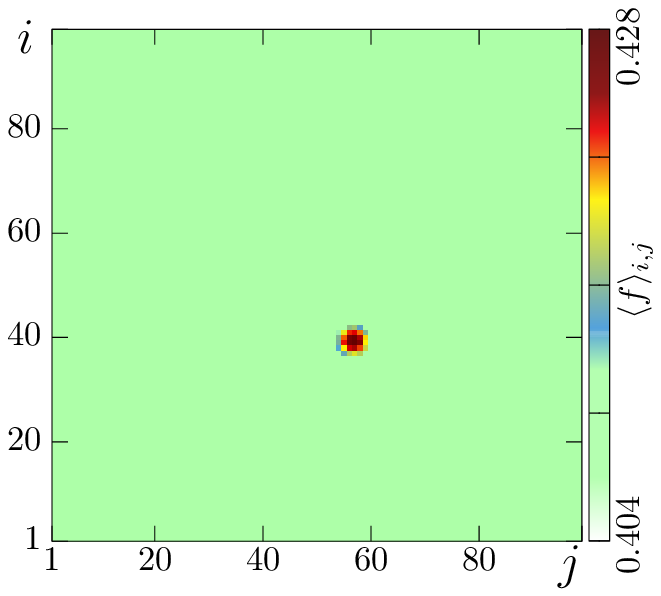}
\vspace{-9.5mm}\center (c) 
}
\parbox[c]{.45\linewidth}{
  \includegraphics[width=\linewidth]{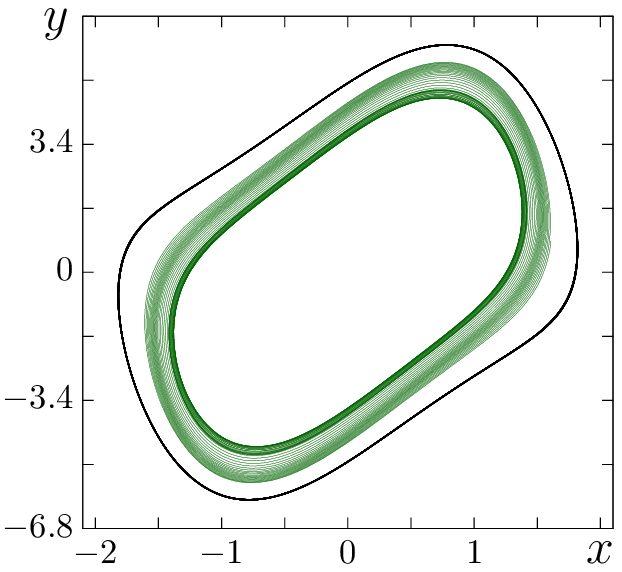}
   \vspace{-9.5mm} \center (b)
\includegraphics[width=\linewidth]{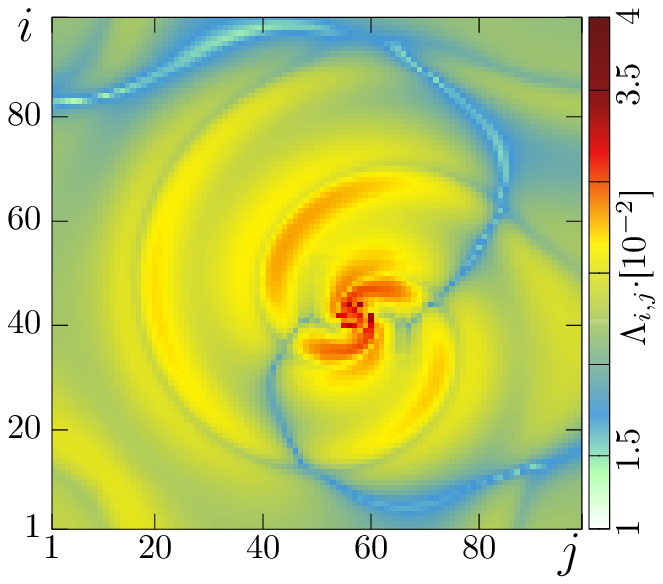}
 \vspace{-9.5mm} \center (d)
}
\caption{(Color online) Spiral wave chimera in \eqref{eq:vdP_grid} for the nonlocal coupling $r=0.09$. (a) is a snapshot of the system state, (b) phase portrait projections for oscillators with indexes $i=44,~j=58 $ (oscillator of the incoherence core, green line) and  $i=40,~j=40 $(synchronous region, black line), (c) spatial distribution of the mean frequency $f^{i,j}$, (d) spatial distribution of the LIS  $\Lambda_{i,j} $. Parameters: $\sigma=0.9 $, $\mu=2.1 $, $\omega=2.5$, $N=100 $}
\label{fig:vdP_P=9}
\end{figure}

We assume that hypersensitivity of the center of a spiral chimera is one of the main reason of the formation of the  incoherence core around the center with increasing in the number of the neighbors coupled with the element. To confirm or refute this hypothesis, we study the evolution of the system for the longer coupling range. At first, we study the case of $r=0.09 $. The stable wave regime for this value of $r $ is represented by a snapshot of the system state in fig.\ref{fig:vdP_P=9}(a).
It is clearly seen that an incoherence core is formed around the wave center, i.e.formation of the spiral wave chimera takes place now. The features of this regime is similar to that of the previous case with ($\lambda_1= 0.020757 $). 
Oscillations corresponding to the incoherence core elements have a chaotic character and correspond to a chaotic attractor, while oscillators outside this core demonstrate the regular dynamics and are characterized by the stable limit cycle. These differences are well visible in projections of the phase portrait of elements of the clusters with the coherent and incoherent behavior, which are shown in fig.\ref{fig:vdP_P=9}(b). Since the oscillator dynamics of the spiral core is chaotic, a value of the maximal Lyapunov exponent for the whole lattice for this regime is equal to $\lambda_1=0.020757$.  It is known that the spiral wave chimera has similar features as in the phase chimera in the ensemble of Kuramoto oscillators, namely, it has the characteristic maximum of the mean frequency distribution in an center of the incoherence core \cite{guo2018spiral, totz2018spiral}. The similar distribution takes place in our case (see fig.\ref{fig:vdP_P=9}(c)). The frequency of the oscillators in the incoherence core smoothly decreases as one moves away from the core center in any radial direction. Fig.\ref{fig:vdP_P=9}(d) shows a spatial distribution of the ILS $\Lambda_{i,j} $.  This confirms that the most sensitive region in the lattice \eqref{eq:vdP_grid} is the core of a spiral wave as well as in the case above. The incoherence core is formed exactly within this spatial region. The most sensitive oscillators are located close to the core boundary, while oscillators inside the incoherence core are characterized by small values of the ILS. Indeed, the oscillations inside the core are noticeably less chaotic than ones in the core boundary.
These results enable us to assume that the reason behind the occurrence of the incoherence core around the spiral wave center is associated with an increase in the sensitivity of oscillators around this center with the elongation of the coupling range $r$. Hence, these elements become more and more unstable and this leads to chaotization of the oscillation process inside the high-sensitive region and as a result formation of the incoherence cluster takes place.

\begin{figure}[!ht]
\centering
\parbox[c]{.45\linewidth}{ 
  \includegraphics[width=\linewidth]{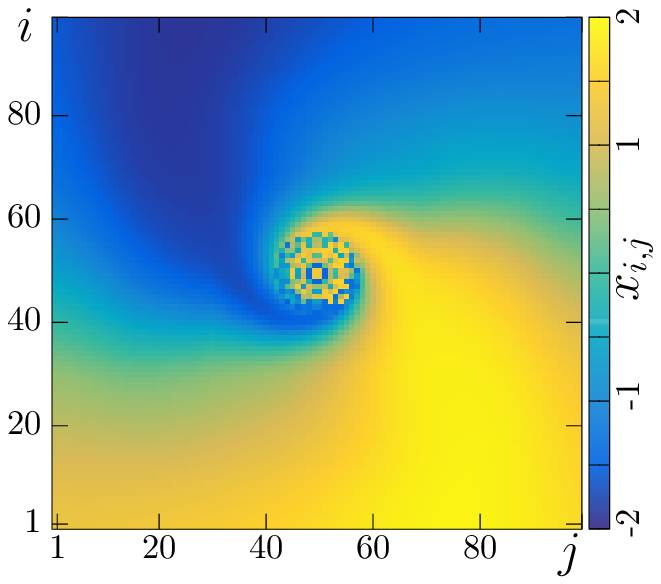}
    \vspace{-9.5mm} \center (a)
}
\parbox[c]{.45\linewidth}{
  \includegraphics[width=\linewidth]{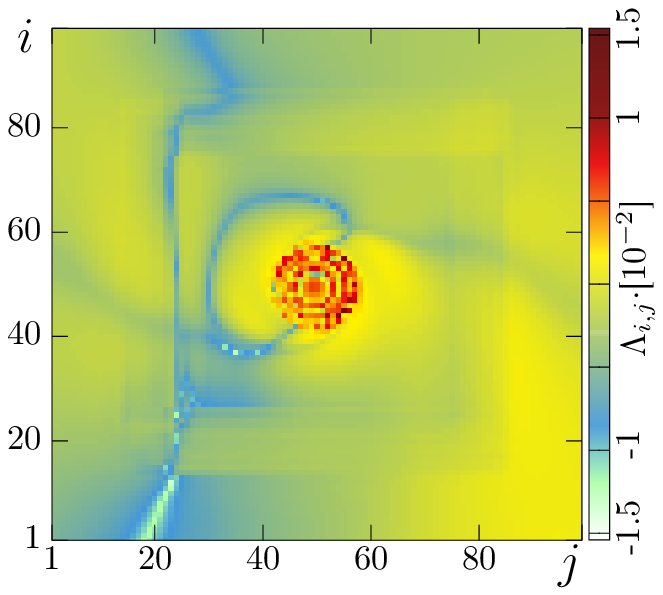}
   \vspace{-9.5mm} \center (b)
}
\caption{(Color online) Spiral wave chimera in \eqref{eq:vdP_grid} for the nonlocal coupling $r=0.3$. (a) is a snapshot of the system state, (b) spatial distribution of the LIS  $\Lambda_{i,j} $. Parameters: $\sigma=0.9 $, $\mu=2.1 $, $\omega=2.5$, $N=100 $}
\label{fig:vdP_P=30}
\end{figure}

Further elongation of the coupling range leads to an increase a number of oscillators forming the incoherence core and accordingly to expansion of this core in space. An example of the spiral wave chimera for the long coupling range $r=0.3 $ is shown in a snapshot of the system state in fig.\ref{fig:vdP_P=30}(a). It should be noted that the spatial configuration of the incoherence core has noticeably changed in comparison with the case of the shorter $r $. Now, the core of the SW chimera has an interesting structure, namely alternating concentric layers of coherent and incoherent oscillators, namely one layer has the incoherent distribution of the instantaneous states, and the following layer is characterized by the synchronous behavior of oscillators.
A spatial distribution of the mean frequency is similar to the previous case and has typical shape for the spiral wave chimeras. Elements of the incoherence core remain the most sensitive in the system and have the maximal values of the ILS, what is shown in fig.\ref{fig:vdP_P=30}(b). Moreover, a spatial structure of the ILS distribution for the incoherence core is also has the ring-like shape, i.e. the ring layers with with high values of the ILS (incoherence layer) alternates with the low-sensitive layers (synchronous layer). 

If we continue to increase a value of the coupling range ($r>0.31 $) then the model \eqref{eq:vdP_grid} switch to the regime of full synchronization when all the lattice elements oscillate periodically with the same instantaneous phases and amplitudes. The frequency of all the oscillators is also the same. Hence, a value of the maximal Lyapunov exponent becomes equal to zero ($\lambda_1=0.000095 $ for $r=0.32 $).

\subsection{FitzHugh-Nagumo model:spiral waves}
Spiral waves in the network of the FHN oscillators regime has been found in \cite{schmidt2017chimera,guo2018spiral,wu2018chimera}. The dynamical regime of the individual oscillators is self-sustained in the system under study \eqref{eq:FHN_grid}. However, the interaction between the oscillators is introduced through a special rotational matrix in all of the above mentioned examples. This type of coupling significantly changes the dynamics of the individual elements of the network. In the system under study, the coupling form between the elements is introduced in the first state variable ($u_{i,j} $ variable, see the eq. \eqref{eq:FHN_grid}). To the best of our knowledge, spiral wave chimeras has not been found in this type of coupling before. Our investigation confirms that formation of this type of chimera takes place in the model  \eqref{eq:FHN_grid}, when the coupling strength $\sigma $ is sufficiently low. Note that the SWC's are observed in the FHN lattice for significantly lower coupling strength ($\sigma \approx 0.08 $ for \eqref{eq:FHN_grid} while they are observed in the VdP lattice for comparatively large coupling strength of $\sigma=0.9 $ for \eqref{eq:vdP_grid}).

\begin{figure}[!ht]
\centering
\parbox[c]{.32\linewidth}{ 
  \includegraphics[width=\linewidth]{vdP-P=1-SnapShot2D}
    \vspace{-9.5mm} \center (a)
}
\parbox[c]{.32\linewidth}{
  \includegraphics[width=\linewidth]{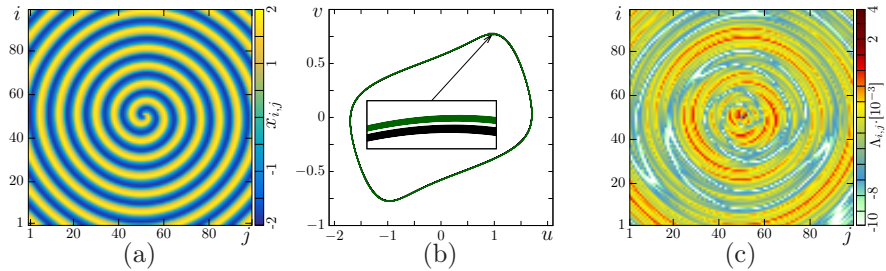}
   \vspace{-9.5mm} \center (b)
}
\parbox[c]{.32\linewidth}{
  \includegraphics[width=\linewidth]{vdP-P=1-LIS}
   \vspace{-9.5mm} \center (c)
}
\caption{(Color online) Spiral wave in \eqref{eq:FHN_grid} for the local coupling $r=0.01$. (a) is a snapshot of the system state, (b) phase portrait projections for oscillators with indexes $i=41,~j=32 $ (wave center oscillator, green line) and  $i=40,~j=40 $(synchronous region, black line), (c) spatial distribution of the LIS  $\Lambda_{i,j} $. Parameters: $\sigma=0.08 $, $\varepsilon = 0.2 $, $\gamma =0.8$, $\beta=0.001 $, $N=100 $}
\label{fig:FHN_P=1}
\end{figure}

 The system is characterized by the high-degree multistability when the coupling is local ($r=0.01 $ or $P=1 $), namely states with a different number of spiral waves can be realized from various randomly distributed initial conditions. We chose the case when only one spiral wave is set in the lattice. This state is illustrated by a snapshot of the system state in  fig.\ref{fig:FHN_P=1}(a). 
Oscillations of all the oscillators including the wave center oscillator have the regular character. A zero value of the maximal Lyapunov exponent ($\lambda_1=0.000196 $) confirms this. The corresponding phase portrait projections are presented in fig.\ref{fig:FHN_P=1}(b).  
At first sight, these attractors correspond to purely periodic oscillations and are the same. However, an enlarged fragment of the trajectories shows that the oscillations poorly express quasi-harmonic character. Furthermore, oscillations in the wave center corresponds to the other attractor than in the other part of the lattice, but very similar. A spatial distribution of the ILS $\Lambda_{i,j} $ is demonstrated in fig.\ref{fig:FHN_P=1}(c). It shows that there is no spatial region with high sensitivity as well as for the spiral chimera in the model \eqref{eq:vdP_grid}. At that instant, the wave center oscillator is the most sensitive element in the whole system. The frequency of this element is also slightly higher than for the other system elements.

\subsection{FitzHugh-Nagumo model: spiral wave chimeras}

\begin{figure}[!ht]
\centering
\parbox[c]{.32\linewidth}{ 
  \includegraphics[width=\linewidth]{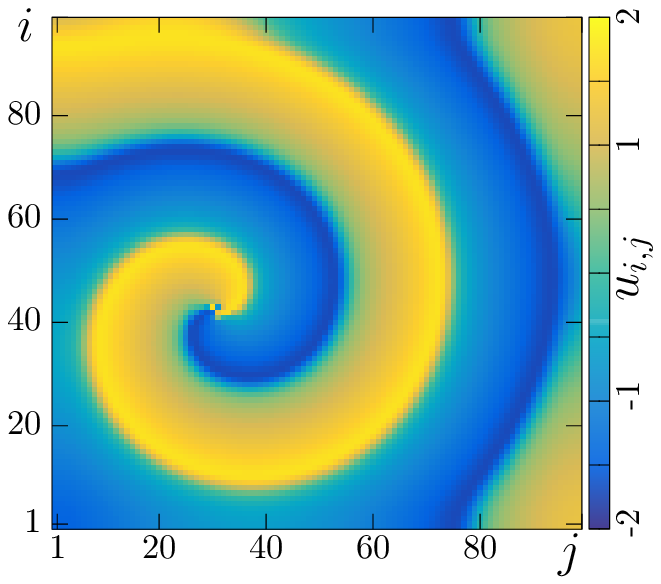}
    \vspace{-9.5mm} \center (a)
}
\parbox[c]{.32\linewidth}{
  \includegraphics[width=\linewidth]{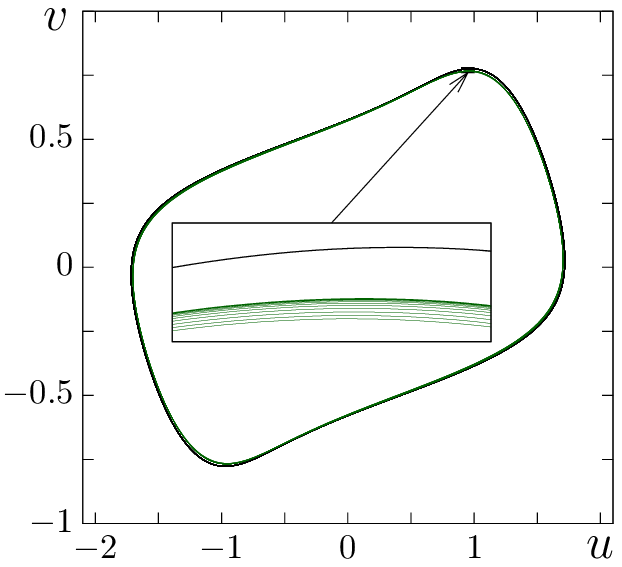}
   \vspace{-9.5mm} \center (b)
}
\parbox[c]{.32\linewidth}{
  \includegraphics[width=\linewidth]{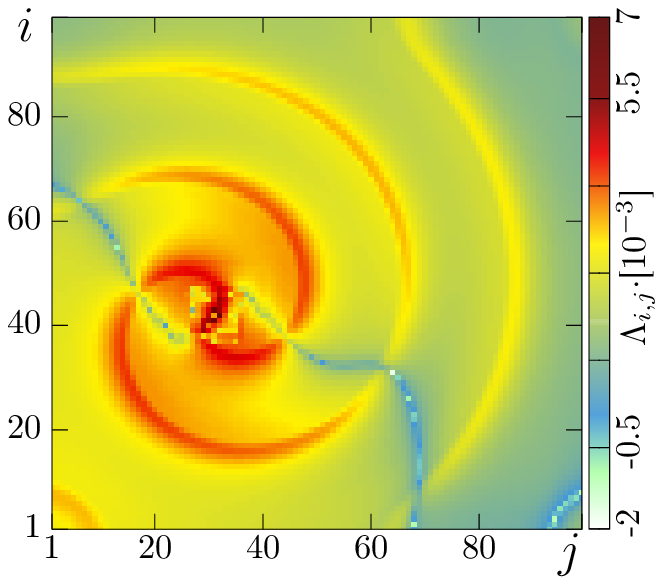}
   \vspace{-9.5mm} \center (c)
}
\caption{(Color online) Spiral wave in \eqref{eq:FHN_grid} for the nonlocal coupling $r=0.04$. (a) is a snapshot of the system state, (b) phase portrait projections for oscillators with indexes $i=43,~j=31 $ (wave center oscillator, green line) and  $i=30,~j=20 $(synchronous region, black line), (c) spatial distribution of the LIS  $\Lambda_{i,j} $. Parameters: $\sigma=0.08 $, $\varepsilon = 0.2 $, $\gamma =0.8$, $\beta=0.001 $, $N=100 $}
\label{fig:FHN_P=4}
\end{figure}

Let us now explore the evolution of the system with an increase in the coupling range values. Growth of the coupling range lead to elongation of the wavelength. The spiral wave realized for the case of $r=0.04 $ is illustrated in fig.\ref{fig:FHN_P=4}(a). The slight incoherence already occurs around the wave center, but the incoherence cluster has not formed yet. Oscillations of the elements which are outside the wave center becomes fully periodic, while oscillations within this center apparently seem to become weakly chaotic,  as evidenced by a low but positive value of the maximal Lyapunov exponent $\lambda_1=0.004283 $. Fig.\ref{fig:FHN_P=4}(b) illustrates the corresponding phase portrait projections. 
A spatial distribution of $\Lambda_{i,j} $ presented in fig.\ref{fig:FHN_P=4}(c) shows that now there is a spatial region around the wave  center  with the highest values of the ILS. Oscillators close to the wave center becomes most sensitive in the system \eqref{eq:FHN_grid}, as well as for the lattice of vdP oscillators \eqref{eq:vdP_grid}. 

\begin{figure}[!ht]
\centering
\parbox[c]{.32\linewidth}{ 
  \includegraphics[width=\linewidth]{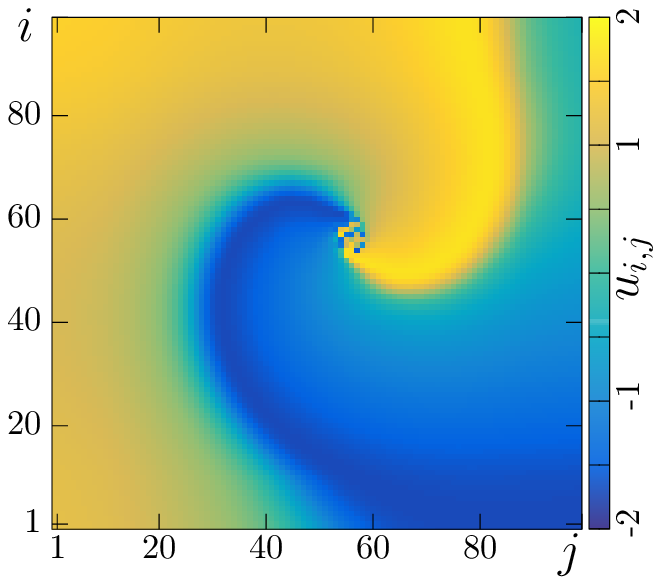}
    \vspace{-9.5mm} \center (a)
}
\parbox[c]{.32\linewidth}{
  \includegraphics[width=\linewidth]{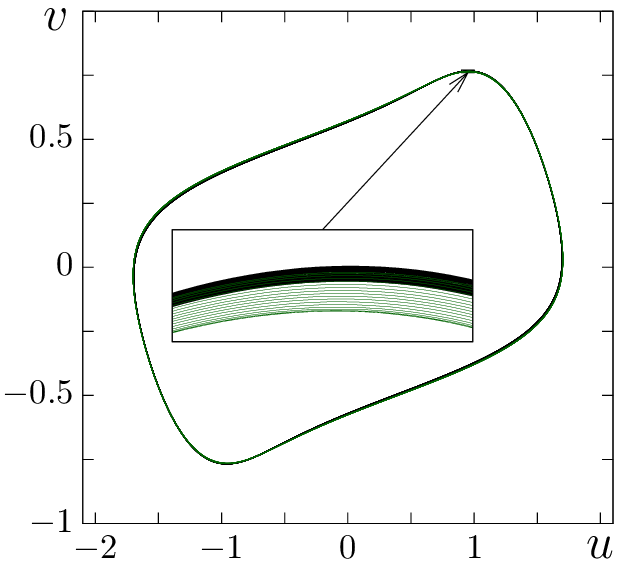}
   \vspace{-9.5mm} \center (b)
}
\parbox[c]{.32\linewidth}{
  \includegraphics[width=\linewidth]{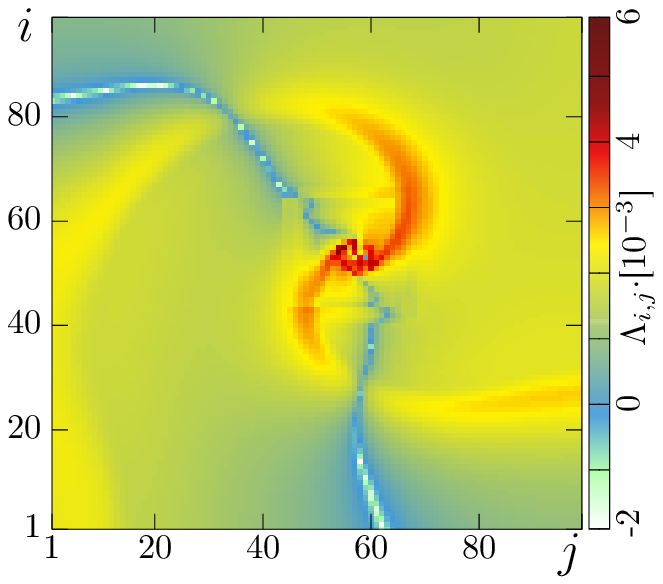}
   \vspace{-9.5mm} \center (c)
}
\caption{(Color online) Spiral wave chimera in \eqref{eq:FHN_grid} for the nonlocal coupling $r=0.1$. (a) is a snapshot of the system state, (b) phase portrait projections for oscillators with indexes $i=54,~j=54 $ (wave center oscillator, green line) and  $i=77,~j=8 $(synchronous region, black line), (c) spatial distribution of the LIS  $\Lambda_{i,j} $. Parameters: $\sigma=0.08 $, $\varepsilon = 0.2 $, $\gamma =0.8$, $\beta=0.001 $, $N=100 $}
\label{fig:FHN_P=10}
\end{figure}

Next we study the wave regime when the coupling range $r $ is extended up to $r=0.1 $. The spiral wave completely transforms to the spiral wave chimera with the presence of an incoherence core. This regime is shown in fig.\ref{fig:FHN_P=10}(a). Oscillations in the incoherence core remain weakly chaotic, which  is seen from the phase portrait projection in fig.\ref{fig:FHN_P=10}(b). The chaos is so weak that the phase portrait projection seems to correspond to a limit cycle. However, a value of the maximal exponent remains positive ($\lambda_1=0.003144 $). Therefore,  oscillations in the lattice should be chaotic. Indeed, if we observe an enlarged fragment of the phase portrait projection then it is possible to see that oscillations in the incoherence core corresponds to a chaotic attractor. Oscillations in the coherence region are also either quasi-periodic or slightly chaotic, but their corresponding  phase portrait projection is significantly more narrow.
Fig.\ref{fig:FHN_P=10}(c) illustrates that the most sensitive oscillators of the system \eqref{eq:FHN_grid} are located in the incoherence core as well as in the previous case. They are characterized by the maximal values of the ILS ($\Lambda_{i,j} $).  It is means that the behavior in this region is less regular and more unstable. Moreover, the oscillators close to the incoherence core boundary are more sensitive than ones in the core center. The similar behavior has been observed for the lattice of vdP oscillators \eqref{eq:vdP_grid} above.

\begin{figure}[!ht]
\centering
\parbox[c]{.45\linewidth}{ 
  \includegraphics[width=\linewidth]{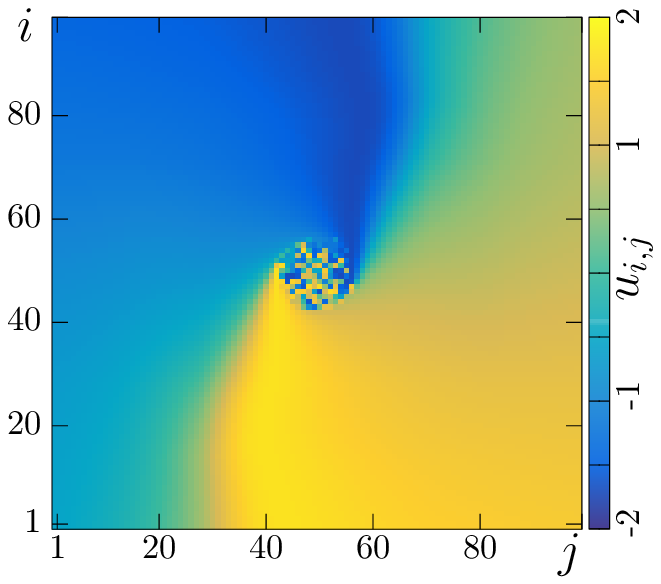}
    \vspace{-9.5mm} \center (a)
}
\parbox[c]{.45\linewidth}{
  \includegraphics[width=\linewidth]{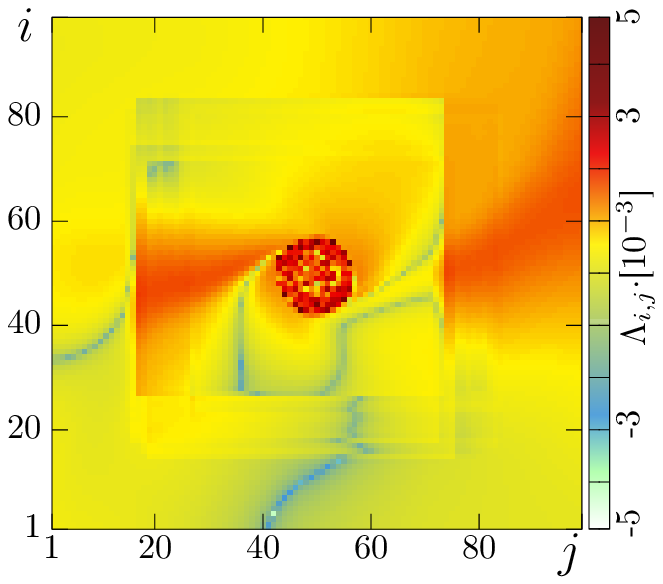}
   \vspace{-9.5mm} \center (b)
}
\caption{(Color online) Spiral wave chimera in \eqref{eq:FHN_grid} for the nonlocal coupling $r=0.28$. (a) is a snapshot of the system state, (b) spatial distribution of the LIS  $\Lambda_{i,j} $. Parameters: $\sigma=0.08 $, $\varepsilon = 0.2 $, $\gamma =0.8$, $\beta=0.001 $, $N=100 $}
\label{fig:FHN_P=28}
\end{figure}

Further increase of the coupling range leads to an extension of the incoherence core and an elongation of the wavelength. An example of the spiral wave chimera in the case of a long coupling range $r=0.28 $ is shown in snapshot of the system state in fig.\ref{fig:FHN_P=28}(a).  
Unlike the SWC in the lattice \eqref{eq:vdP_grid}, the incoherence core in the present case has a more common spatial structure, namely the incoherence is observed within the whole incoherence cluster. A value of the maximal Lyapunov exponent becomes noticeably lower than for shorter $r $ and is equal to $\lambda_1= 0.001232 $, hence the oscillations in the incoherence core are very weakly chaotic. 
A spatial distribution of the indexes of local sensitivity $
\Lambda_{i,j} $ is shown in fig.\ref{fig:FHN_P=28}(b). The elements with characterizing by the maximal values of $
\Lambda_{i,j} $ are located in the incoherence cluster boundary, while oscillators inside the cluster are less sensitive. The same character of the ILS distribution is observed for the shorter coupling ranges. A spatial distribution of the frequency in the incoherence core is also have a typical bell-like shape.

\subsection{Comparison}

Comparing the evolution of the spiral wave regime in the lattice of van der Pol oscillators \eqref{eq:vdP_grid} and that of the FHN oscillators \eqref{eq:FHN_grid}, we identify a similar scenario, which is observed for both systems. When the interaction between elements has the local character, the wave center sensitivity does not differ from the rest part of the system. All the oscillators demonstrate the regular behavior. However, growth of the coupling range values leads to a sharp increase in the sensitivity of a spatial region around the wave center. Oscillations inside the center become chaotic, while they remain regular outside the center core. Furthermore, this region extends in space. As a result, the incoherence appears inside this region and incoherence cluster forms as the wave core, when the coupling range $r $ becomes sufficiently long. This core extends with growth of $r $ up to sufficiently large values of $r \approx 0.3 $. When the value of the coupling range exceeds this threshold level, both models switch to the regime of complete synchronization, when all the elements oscillate periodically as a whole with the same instantaneous amplitudes and phases.

At that instant, the dynamics of the two models under study is not the same. The lattice of vdP oscillators demonstrates more complex dynamics. Oscillations in the incoherence cluster are characterized by more expressed chaotic behavior than for the case of the FHN model, where oscillations is almost regular. Apparently, this is the reason why the incoherence core in the model \eqref{eq:vdP_grid} has a complex spatial structure for the long coupling ranges.

\subsection{Comparison of evolution of the Lyapunov spectrum for the varied coupling range.}

\begin{figure}[!ht]
\centering
  \includegraphics[width=0.72\linewidth]{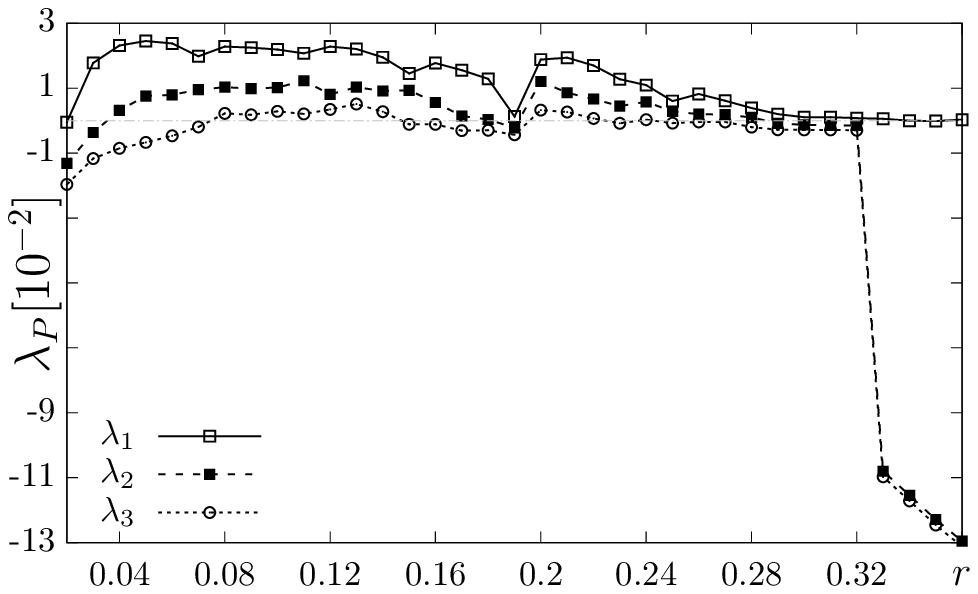}\\
    \vspace{-4.5mm} 
    \center (a)\\
  \includegraphics[width=0.72\linewidth]{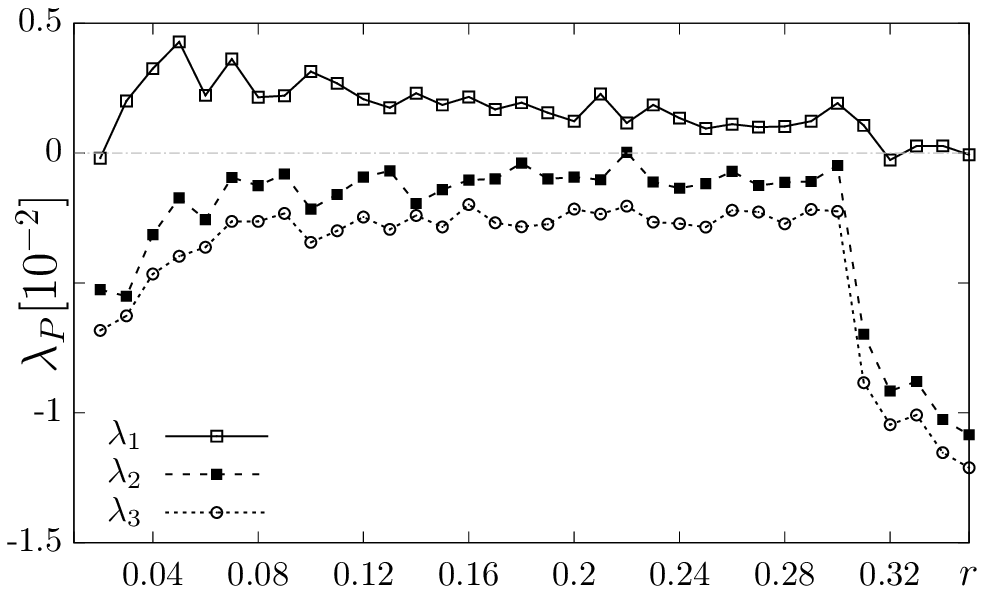}\\
   \vspace{-4.5mm} \center (b)
\caption{(Color inline) Dependences of the first three Lyapunov exponents $\lambda_1 $, $\lambda_2 $ and $\lambda_3 $ on the coupling range $r $ for the systems  \eqref{eq:vdP_grid} (a) and \eqref{eq:FHN_grid} (b). Parameters of \eqref{eq:vdP_grid}: $\sigma=0.9 $, $\mu=2.1 $, $\omega=2.5$, $N=100 $; parameters of \eqref{eq:FHN_grid}:$\sigma=0.08 $, $\varepsilon = 0.2 $, $\gamma =0.8$, $\beta=0.001 $, $N=100 $}
\label{fig:Lyapunov}
\end{figure}

It is interesting to compare the evolution of the Lyapunov spectrum for both systems under study when the number of coupled neighbors increases. Are the systems in a state of the hyperchaos (i.e. has more than one positive Lyapunov exponents (LE) $\lambda_k$)? To answer these questions, we calculate the evolution of the first three Lyapunov exponents for both of the lattice \eqref{eq:vdP_grid} and \eqref{eq:FHN_grid}, when values of the coupling range $ r $ increase. At first we explore the dependence for the vdP model \eqref{eq:vdP_grid} which is presented in fig.\ref{fig:Lyapunov}(a). When the coupling is local, the system demonstrates the regular behavior with the zero maximal LE and negative for others. A sharp switch of the maximal LE to positive value region is observed already for $r=0.02 $, but for others LE's remain negative. However, the system switches to the hyperchaotic regime beginning with a value of $r \geq 0.03 $ (the second LE becomes positive while the third LE remains negative). The maximal LE reaches the maximum value when the coupling range becomes equal to $r=0.04 $ and decreases for longer $r$. A sharp increase in the sensitivity of a region around the wave center is observed exactly for this value of coupling range (the oscillator dynamics around the wave center becomes strongly chaotic, while the remaining part of the system remains to oscillate regularly). The third LE exceeds a zero value for $r=0.07 $.
The special behavior of the model \eqref{eq:vdP_grid} is observed when the coupling range approaches to the value $r=0.18$. The values of all the LE's decreases. At that instant, the size of the incoherence cluster shrinks, and a region with a smooth spatial profile occurs instead of the incoherence. At $r=0.18 $, all the three LE's reach the minimal values within $r \in [0.1;0.3] $, namely $\lambda_1=0.001251 $ and the other LE's become negative. The spatiotemporal structure undergoes significant change for this value of $r $. The incoherence cluster completely disappears and a spatial region with the synchronous dynamics begins to form instead of the incoherence core. Further increase in the the coupling range leads to an occurrence of the incoherence again and growth of values of all the three first LE's takes place. However, when values of $r $ exceed $0.32 $, the regime of spiral wave chimeras are destroyed and the system switches to the complete synchronization regime with purely regular dynamics of the oscillators. At that we observe a sharp decline of the 2nd and 3rd LE's to negative values while the maximal LE $\lambda_1 $  becomes equal to zero within the calculation accuracy. 

The dynamics of the second model \eqref{eq:FHN_grid} is simpler in comparison with the lattice \eqref{eq:vdP_grid}. The evolution of $\lambda_{1,2,3} $ is exemplified in fig.\ref{fig:Lyapunov}(b). When the coupling is local ($r=0.01 $), the maximal LE $\lambda_1 $ is equal to zero and the other LE's are negative. But  $\lambda_1 $ becomes positive for $r=0.02 $ and reaches the maximal value $\lambda_1=0.004283 $, when the coupling range is equal to $r=0.04 $. Values of $\lambda_{1,2} $ are maximal for this $r $.  The spatiotemporal incoherence also appears in the FHN model within the high-sensitive area around the wave center, which forms when the coupling range is elongated. 
It should be noted that at the maximal value of $\lambda_1 $ is significantly less than that of the van der Pol model.
Moreover, the lattice \eqref{eq:FHN_grid} is never switched to the hyperchaotic regime, because values of $\lambda_{1,2} $ always remain negative. The second LE reaches a zero value at $r=0.22 $, but latter does not lead to any qualitative changes of the SWC. The FHN model switches to the regime of complete synchronization (as well as the vdP model) when $r $ exceeds the value $r>0.3 $. The maximal LE becomes equal to 0 while the values of $\lambda_{2,3} $ sharply decreases. 
These results approve that the dynamics of the lattice \eqref{eq:FHN_grid} in the spiral wave chimera regime is characterized by a noticeable less chaotic dynamics than in the vdP model, which demonstrates the more complex behavior. Hence, the dependency of the maximal LE on the coupling range has a similar character for both models. Hence, it enables us to conclude that formation mechanisms of SW chimera has the same nature.

Thus, evolution of the maximal Lyapunov exponent and of the ILS distribution enables us to simply diagnose the transformation of the spiral wave to the spiral wave chimera. Elongation of the coupling interval leads to a loss of the stability of a spiral wave core (center) (ILS values are maximal for this spatial region). As a result, the type of oscillator dynamics within the wave center changes from regular to chaotic (or even hyperchaotic) one. At that instant, other oscillators of the lattice continue to oscillate almost periodically. In consequence of these processes, the incoherence cluster forms around the spiral wave core. Switching of the models to the synchronous regime is accompanied by sharp decline of the second and third LE's.

\section*{Conclusions}

We have studied the evolution of the auto-wave regimes by increasing the nonlocality degree of two different networks connected in two-dimensional geometry with the free flux boundary conditions. The first network consists of the van der Pol oscillators and the second one is of the FitzHugh-Nagumo (FHN) oscillators with self-sustained dynamics. The main auto-wave regime for these systems is a spiral wave, which transforms into a spiral wave chimera with the growth of the nonlocality degree. The features of the spiral wave chimeras under study are similar to the ones in  the other models \cite{bukh2019bspiral,kuramoto2003rotating,tang2014novel,li2016spiral,shima2004rotating,kundu2018diffusion}. For example, there is no characteristic bell-like spatial distribution of the mean oscillation frequency. This enables us to assume that the formation of  incoherence core around the spiral wave center has the same features.
Thus, the aim of the current study is concentrated on search and explanation of some common quantitative peculiarities of the formation of the spiral wave chimera from the spiral wave. 

The spiral waves appear in both models under study under the local coupling and also exists for low non-locality degree. However, in the van der Pol model, they are observed for a significantly higher coupling strength ($\sigma \approx 0.9 $) than that of the FHN model ($\sigma \approx 0.08 $). Our comparative study has demonstrated that, eventhough the dynamics of a single oscillator in both of the models are described by different dynamical equations, both systems demonstrate a similar behavior when the non-locality degree increases. The dynamics of all the individual oscillators of both systems is regular (periodic or quasi-periodic), which is confirmed by a zero value of the maximal Lyapunov exponent (LE). But even a small elongation of the coupling range (non-locality degree) leads to switching of the oscillations of the spiral wave center from the regular to chaotic, while the dynamics of the remaining part of the system remains regular. Introducing non-local coupling induces a change in the lattice. We have calculated the indices of local sensitivity (ILS) described in \cite{shepelev2018local} for the quantitative analysis of the spatiotemporal auto-wave structures. The spatial distribution of ILS enables us to evaluate the most and least sensitive spatial regions in the structures.  
When the coupling is local, the sensitivity is similar for the whole system. The introduction of the nonlocal coupling leads to the formation of a highly sensitive spatial region around the spiral wave center. Moreover, this region extends with the growth of the nonlocal coupling. Values of the maximal LE  also increase up to a certain threshold value of the coupling range (sufficiently short), when it reaches the maximum value. 
Chaotic oscillations are observed only close to the wave center. This means that these oscillations are most chaotic for the threshold level of the coupling range, but there is not yet any incoherence in the lattices. Apparently, for this reason, oscillators stop to demonstrate the synchronous behavior, when values of the coupling range exceed this level, and the incoherence core begins to form around the wave center. The sensitivity of this core remains maximal in the lattice. The size of the core extends with the elongation of the coupling range. It should be noted that the spatiotemporal dynamics of the lattice of van der Pol oscillators are more complex than that of the FHN model. The formation of the incoherence core is accompanied by switching of the system to hyperchaotic regime, where several of the Lyapunov exponents are positive, while the regime with only one positive LE is observed in the second system under study. Indeed, the chaotic behavior of the oscillators in the incoherence core are weakly expressed by this model. 

Thus, we show that the features of formation of the incoherence core in the lattice of oscillators under study are following: Introducing the non-locality into the coupling leads to the formation of an instability region with increasing sensitivity around the spiral wave center. When the non-locality exceeds a certain threshold level, the elements near the centre of the wave become so unstable that they cease to oscillate synchronously with each other and form the incoherence core. The size of the highly-sensitive spatial region  with chaotic dynamics increases with the elongation of the coupling range. The incoherence core accordingly expands too. A similar behavior has been discovered in the lattice of FHN oscillators in the bistable regime in \cite{shepelev2019variety} and for the lattice of discrete-time oscillator in the appendix.

\section{Appendix}

We also study the evolution of a spiral wave when the coupling range increases for the lattice of discrete-time oscillators (maps). A basic element of this lattice is the Nekorkin map described as following:
\begin{equation}\label{eq1-map}
\begin{aligned}
& x^{t+1}= x^t + F(x^t) - y^t - \beta H (x^t - d),\\
& y^{t+1}= y^t + \varepsilon (x^t - J),
\end{aligned}\end{equation}
where $x^t$ is a variable that describes the dynamics of the membrane
potential of the nerve cell, $y^t$ is a variable that relates to the
cumulative effect of all ion currents across the membrane, functions
$F(x^t)$ and $H(x^t-d)$ are given as follows:
\begin{equation}\label{eq2-F}
\begin{aligned}
& F(x^t) = x^t(x^t-a)(1-x^t), \qquad 0<a<1,
\\
& H(x^t) = 
\begin{cases}
1, \qquad x^t>0,	\\
0, \qquad \mbox{elsewhere}.
\end{cases}
\end{aligned}\end{equation}
The parameter $\varepsilon>0$ determines the characteristic time scale of
$y^t$, the parameter $J$ controls the level of the membrane depolarization
$(J<d)$, the parameters $\beta>0$ and $d>0$ determine the excitation
threshold of bursting oscillations, $t=1,2,\ldots$ represents the discrete
time. 

A $N\times N$ 2D lattice of the nonlocally coupled Nekorkin maps is described by the following system of equations:
\begin{equation}\label{eq:Nek-lattice}
\begin{aligned}
x_{i,j}^{t+1}&= x_{i,j}^t + F(x_{i,j}^t) - y_{i,j}^t - \beta H (x_{i,j}^t - d) +,\\
& + \dfrac{\sigma_x}{B_{i,j}^x} \sum\limits_{m_x,n_x} \left[ f(x_{m_x,n_x}^t) - f(x_{i,j}^t) \right],\\
y_{i,j}^{t+1}&= y_{i,j}^t + \varepsilon (x_{i,j}^t - J),
\end{aligned}
\end{equation}
where $m_x,n_x\in\mathbb{N}$ are indices of the nonlocal neighbours. The sum denotes the nonlocal coupling of range $R_x$ in a square domain. The
parameter $\sigma_x$ denotes the coupling strength between the elements in the $x$ variable, $B_{i,j}^x$ gives the number of nonlocally coupled
neighbors of node ($i,j$).

\begin{figure}[!ht]
\centering
\parbox[c]{.45\linewidth}{ 
  \includegraphics[width=\linewidth]{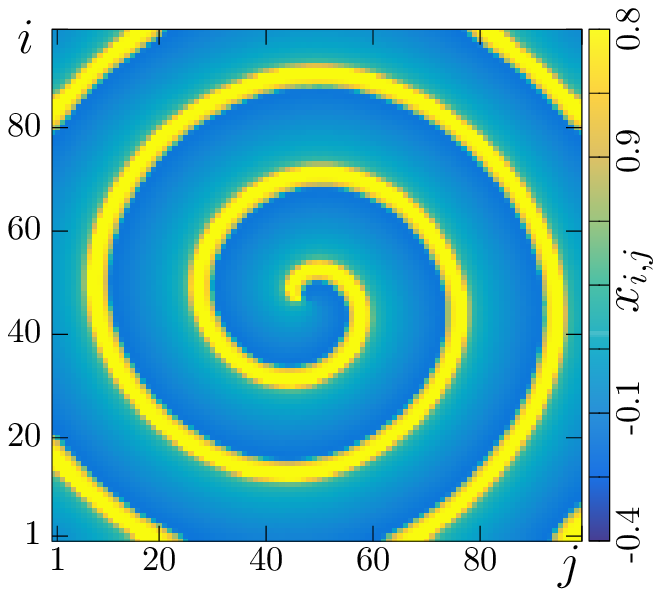}
    \vspace{-9.5mm} \center (a)
\includegraphics[width=\linewidth]{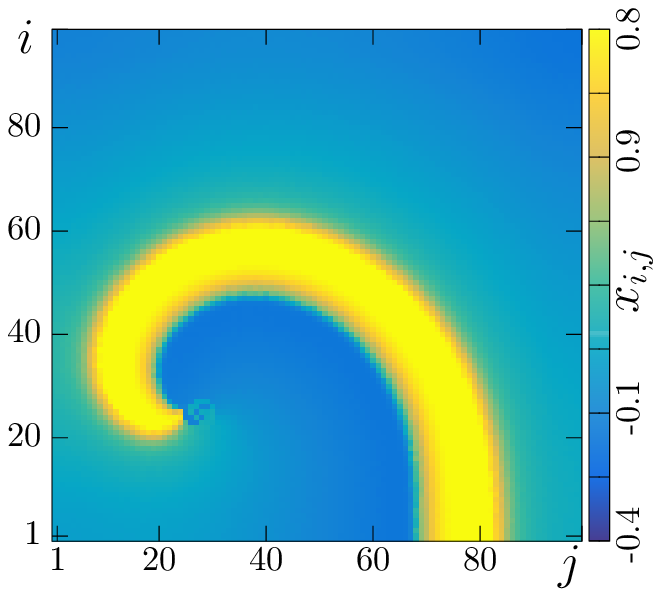}
\vspace{-9.5mm}\center (c) 
\includegraphics[width=\linewidth]{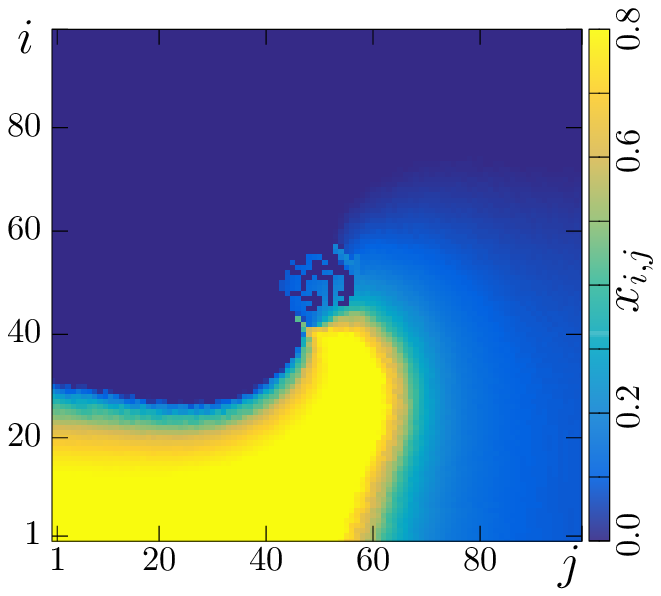}
\vspace{-9.5mm}\center (e) 
\includegraphics[width=\linewidth]{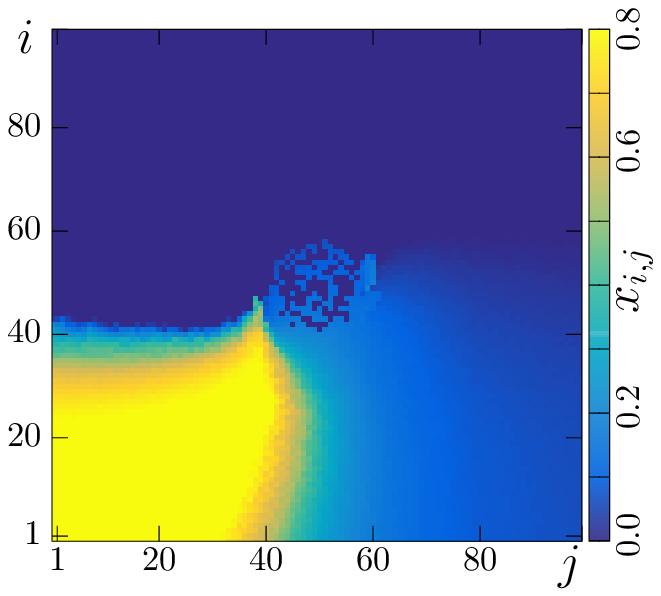}
\vspace{-9.5mm}\center (g) 
}
\parbox[c]{.45\linewidth}{
  \includegraphics[width=\linewidth]{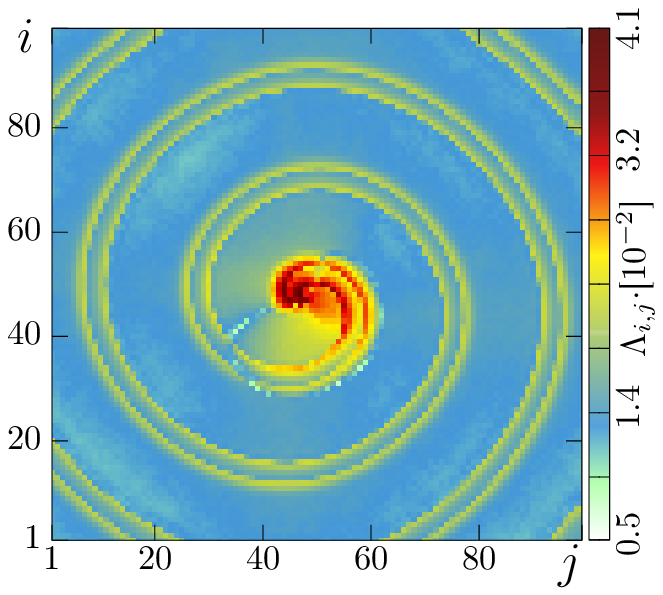}
    \vspace{-9.5mm} \center (b)
\includegraphics[width=\linewidth]{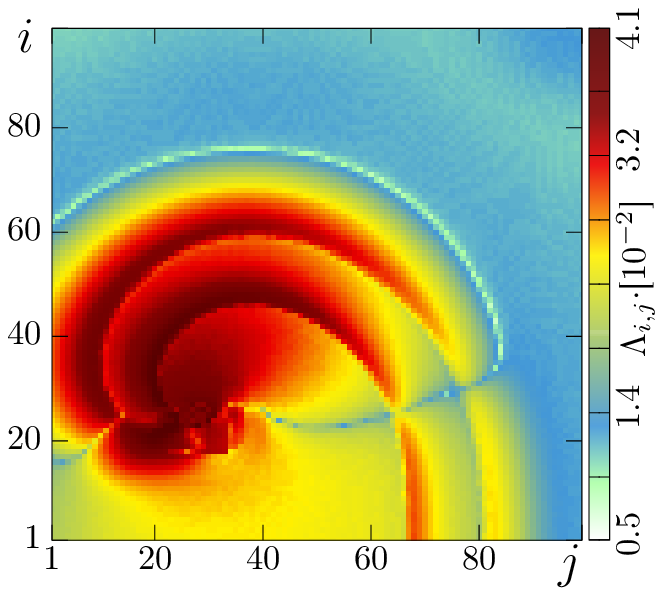}
\vspace{-9.5mm}\center (d) 
\includegraphics[width=\linewidth]{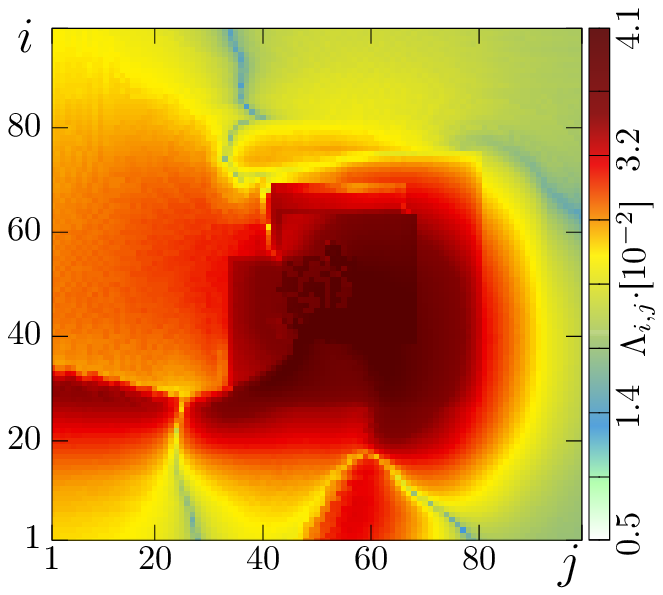}
\vspace{-9.5mm}\center (f) 
\includegraphics[width=\linewidth]{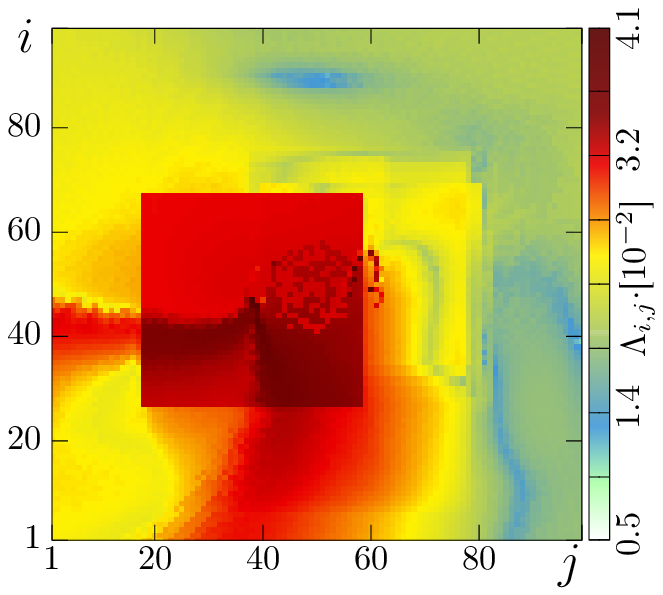}
\vspace{-9.5mm}\center (h) 
}
\caption{(Color online) Spiral waves and spiral wave chimeras in \eqref{eq:Nek-lattice} for $\sigma_x=0.05$. The left column illustrates snapshots of the system states and the right column shows spatial distributions of the ILS, when coupling radius equal to (a,b) $r=1$, (c,d) $r=5$, (e,f) $r=12$, (g,h) $r=20$. Parameters: $J=0.15$, $\varepsilon=0.004$, $\alpha=0.25$, $\beta=0.04$, $d=0.5$, $N=100 $ }
\label{fig:Nek}
\end{figure}
\begin{figure}[!ht]
\centering
  \includegraphics[width=0.7\linewidth]{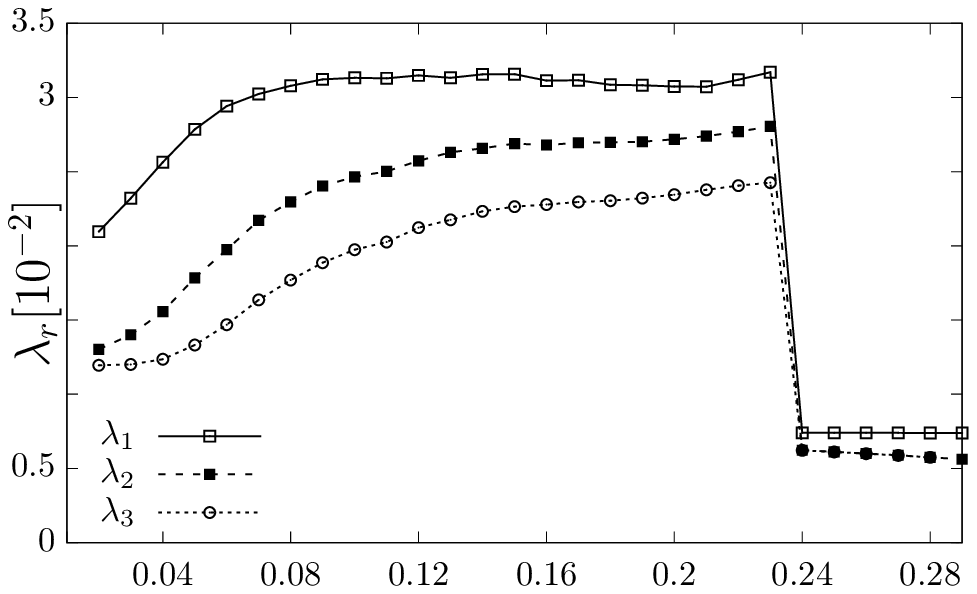}\\
    \vspace{-4.5mm} 
\caption{(Color inline) Dependences of the three first Lyapunov exponents $\lambda_1 $, $\lambda_2 $ and $\lambda_3 $ on the coupling range $r $ for the systems  \eqref{eq:Nek-lattice}. Parameters: $\sigma_x=0.05$, $J=0.15$, $\varepsilon=0.004$, $\alpha=0.25$, $\beta=0.04$, $d=0.5$, $N=100 $}
\label{fig:Lyapunov_Nek}
\end{figure}

The numerical results show that when the nonlocal coupling strength  $\sigma_x$ and the coupling
range $R_x$ are varied, the model~\eqref{eq:Nek-lattice} can demonstrate all
the typical spiral wave patterns, including spiral wave chimeras, which
were observed earlier. Examples of these states are presented from the left in fig.\ref{fig:Nek}. It is seen that the evolution of a spiral wave with increasing values of the coupling range has the same character as that of the systems described above. However, the region around the wave center has high sensitivity even when the coupling is local. This happens as the wave center element oscillates chaotically. Even a slight elongation of the coupling range $r $ leads to significant expansion of the high-sensitive region around the wave center in space. The formation of the incoherence core has already taken place for $r=0.05 $. The right column in fig.\ref{fig:Nek} illustrates the spatial distributions of the ILS with the growth of the coupling strength.
The evolution of the first three LEs for the system \eqref{eq:Nek-lattice} is demonstrated in fig.\ref{fig:Lyapunov_Nek}. The behavior of all the LEs is simpler than that of the models \eqref{eq:vdP_grid} and \eqref{eq:FHN_grid}.
The lattice is always in hyperchaotic regime (even for the very short and very long coupling ranges) because all the three calculated LE's are positive. The coupling range elongation leads to growth of the LE's values. The value of $\lambda_1 (r) $ at first monotonically increases and after that almost does not change with increase in the coupling range. The value of $\lambda_2 (r)$ and $\lambda_3(r) $ also demonstrate monotonous growth. When the model \eqref{eq:Nek-lattice} switches to the synchronous regime at $r=0.23 $ values of all the three LEs sharply decrease, but remain positive.

\section*{Acknowledgements}
\vspace{0.5cm}
This work was funded by the Deutsche Forschungsgemeinschaft (DFG, German Research Foundation) -- Projektnummer. 163436311-SFB 910. I.A.S., A.V.B. and V.S.A.  thank for the financial support provided by RFBR and DFG according to the research project \#20-52-12004,  S.S.M. acknowledges the use of New Zealand eScience Infrastructure (NeSI) high performance computing facilities as part of this research.

\bibliographystyle{plain}
\bibliography{SWC_quantify}

\end{document}